\documentclass{aa} 

\usepackage{graphicx}
\usepackage{txfonts}

\begin{document}

\title{Tracking the motion of a shock along a channel in the low solar
corona}

   \author{J. Rigney\inst{1}\fnmsep\inst{2}\fnmsep\inst{3}\thanks{E-mail: jeremy.rigney@dias.ie}, 
   P.~T.~Gallagher\inst{1}, 
   G.~Ramsay\inst{2}, 
   J.~G.~Doyle\inst{2},
   D.~M.~Long \inst{4,3},
   O. Stepanyuk\inst{5}, 
   \and
   K.~Kozarev \inst{5}}
   
\authorrunning{Rigney et al.}
\titlerunning{Shock motion tracking in the solar corona}

\institute{Astronomy \& Astrophysics Section, DIAS Dunsink Observatory, Dublin Institute for Advanced Studies, Dublin, D15 XR2R, Ireland
      \and
          Armagh Observatory and Planetarium, College Hill, Armagh BT61 9DG, N. Ireland
      \and
          School of Mathematics and Physics, Queen’s University Belfast, University Road, Belfast BT7 1NN, N. Ireland
      \and
          Centre for Astrophysics and Relativity, School of Physical Sciences, Dublin City University, Dublin, D09 V209, Ireland
      \and 
          Institute of Astronomy and National Astronomical Observatory, Bulgarian Academy of Sciences, Tsarigradsko Chausee Blvd 72, Sofia 1784, Bulgaria
}
 
\date{Received 31 October, 2023; }

\abstract
    {Shock waves are excited by coronal mass ejections (CMEs) and large-scale extreme-ultraviolet (EUV) wave fronts and can result in low-frequency radio emission under certain coronal conditions.}    
    {In this work, we investigate a moving source of low-frequency radio emission as a CME and an associated EUV wave front move along a channel of a lower density, magnetic field, and Alfv{\'e}n speed in the solar corona.} 
    {Observations from the Atmospheric Imaging Assembly on board the Solar Dynamics Observatory, the Nançay Radio Heliograph (NRH), and the Irish Low Frequency Array (I-LOFAR) were analysed. Differential emission measure maps were generated to determine densities and Alfv{\'e}n maps, and the kinematics of the EUV wave front was tracked using CorPITA. The radio sources' positions and velocity were calculated from NRH images and I-LOFAR dynamic spectra.}
    {The EUV wave expanded radially with a uniform velocity of $\sim$~500~km~s$^{-1}$. However, the radio source was observed to be deflected and appeared to move along a channel of a lower Alfv{\'e}n speed, abruptly slowing from 1700~km~s$^{-1}$ to 250~km~s$^{-1}$ as it entered a quiet-Sun region. A shock wave with an apparent radial velocity of $>$~420~km~s$^{-1}$ was determined from the drift rate of the associated Type II radio burst.}
    {The apparent motion of the radio source may have resulted from a wave front moving along a coronal wave guide or by different points along the wave front emitting at locations with favourable conditions for shock formation.}


    \maketitle


\section{Introduction}

Observations of solar activity such as flares and coronal mass ejections (CMEs) can help us to better understand magnetic field dynamics, particle acceleration, and plasma motion. A more complete picture of these events can be obtained through multi-wavelength observations \citep[cf.][]{warmuth2004p1,raftery2009}. Extreme-ultraviolet (EUV) observations from instruments, such as the Atmospheric Image Assembly \citep[AIA;][]{lemen2012} on board the Solar Dynamics Observatory \citep[SDO;][]{Pesnell:2012}, image the upper chromosphere and corona, detecting events such as flares, CMEs, and EUV waves \citep{liu2014}. EUV waves are defined as 'arc-shaped fronts' that propagate through the low corona from a shock origin and appear as the footprint of the outward propagating shock front \citep{gallagher2011,Long2011,Long:2017}. Measuring the velocity of EUV waves can give us insight into plasma conditions in the low solar corona and the dynamics of plasma motion in high-strength magnetic field environments \citep{gallagher2003,Long:2013}. While their origin is still disputed, a clear association has been made between the formation of EUV waves at the base of shocks \citep{Thompson1998, Long:2019}. EUV waves are also known to undergo reflection and refraction at coronal hole boundaries, regions of distinct change in the plasma density, and magnetic field environments \citep{gopalswamy2009}.

Low-frequency solar radio emission ($<$~1~GHz) is emitted by accelerated electrons in the corona \citep{wild1950}. Type II radio bursts are formed when accelerated plasma surpasses the local Alfv\'en speed in the surrounding undisturbed environment to form a shock \citep{nelsonmelrose1985}. These super-Alfv\'enic shocks typically occur higher in the solar corona in lower-density plasma environments where the Alfv\'en speed is lower. In dynamic spectra, Type II bursts are characterised by fundamental and harmonic bands that drift from high to low frequency over time, corresponding to the changing plasma frequency at different heights in the corona. Fine structures within these bands reveal electron acceleration and conditions within the shock \citep{Magdalenic2020}. Solar radio imaging during these events can also help determine the propagation of the shock front and regions of electron acceleration. Instruments such as the LOw Frequency ARray \citep[LOFAR;][]{vanHaarlem2013} and the Nançay Radio Heliograph \citep[NRH;][]{kerdraon2007} are capable of providing high-resolution imaging and dynamic spectra of the Sun via dedicated monitoring programmes.

\begin{figure*}
   \centering
   \includegraphics[width=0.98\textwidth]{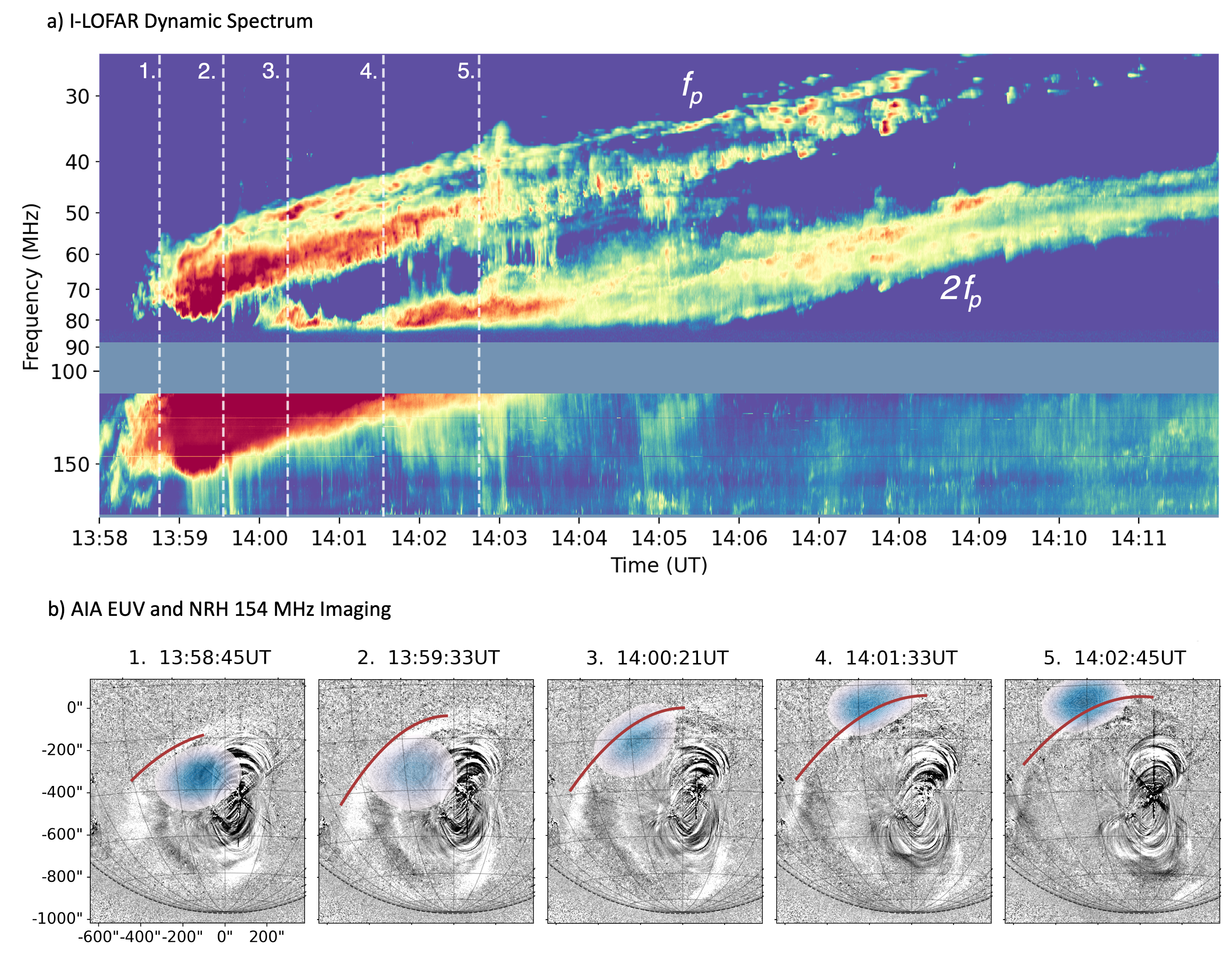}
   
   \caption{ Summary of 10 May 2022 event captured by three instruments. (a): I--LOFAR dynamic spectrum of the Type II radio burst, highlighting the bright fundamental and harmonic bands. The fundamental band drifts from $\sim$~60~MHz to $\sim$~25~MHz over 15~min, with band-splitting occurring within both the fundamental and harmonic lanes. Dashed vertical lines correspond to the timestamps of images in the lower panel. (b): AIA three--channel difference imaging. Overlaid in blue in each panel is the $3\sigma$--to--peak NRH 154 MHz radio source. The solid red line is the EUV wave front that was identified from the CorPITA  difference imaging.}
   \label{fig:1_dynspec_euv}
\end{figure*}

The relationship between Type II radio bursts and EUV waves has been studied extensively \citep{vrsnak2005, magdalenic2008, kouloumvakos2014, morosan2019}. Prior to the first observation of global EUV waves in 1997 \citep{Dere:1997, Thompson1998}, Type II radio bursts were thought to be produced by the same physical process driving Moreton-Ramsey waves \citep[first observed by][]{Moreton:1960a,Moreton:1960b}. However, the significant discrepancy in velocity between Moreton-Ramsey and EUV waves complicated this assumption \citep[see, e.g. the reviews of EUV wave observations by][]{gallagher2011,Warmuth:2015}, leading to a series of statistical studies examining their relationship. Initially, \citet{klassen2000} found that 90\% of Type II radio bursts observed in 1997 were associated with global EUV waves. However, subsequent work by \citet{Biesecker:2002} found that only 29\% of the EUV waves in the catalogue compiled from SOHO/EIT observations by \citet{Thompson:2009} had associated Type II radio bursts. \citet{Muhr:2014} obtained comparable results, finding that 22\% of 60 EUV waves observed using the higher-resolution imaging provided by STEREO/EUVI had an associated Type II radio burst. The advent of very high-resolution observations from SDO/AIA enabled \citet{Nitta:2013} to assemble and examine a catalogue of 138 global EUV waves, finding that 54\% were associated with Type II radio bursts. However, using the catalogue of \citet{Nitta:2013} as a starting point, \citet{long2017} found no clear relationship between global EUV waves identified using the automated Coronal Pulse Identification and Tracking Algorithm \citep[CorPITA;][]{Long:2014} and Type II radio bursts. The relationship between these phenomena therefore continues to be a source of much investigation, with \citet{Fulara:2021} suggesting that Type II bursts are more related to the velocity of the erupting CME than the laterally propagating global EUV wave. 

Most observations capture events on or near the limb, as the height and outward propagation of radio emission can be more accurately determined. Previous research with similar observations of EUV waves and Type II radio bursts have, for the most part, concluded that these events are linked via shock formation from flares and CMEs (see \citet{grechnev2012}). While much of the previous research has examined the links between EUV waves and radio emission, it is difficult to capture high-cadence multi-wavelength (X-ray, EUV, and radio) observations for a single event containing a high-energy flare, shock, and EUV wave. Furthermore, on-disk events can be difficult to characterise but are important for space weather as rapid on-disk events can release sudden energetic particle (SEP) events towards Earth \citep[e.g.][]{desai2016}.

In this Letter observations of an impulsive X-class flare from 10 May 2022 were presented, where both a shock front and EUV wave were observed by SDO and the Geostationary Operational Environmental Satellite (GOES), Irish LOw Frequency ARray \citep[I-LOFAR][]{Murphy2021}, the Observations Radio pour Fedome et l’Etude des Eruptions Solaires radio-spectrograph \citep[ORFEES;][]{hamini2021}, and the NRH. We summarise the characteristics of the radio emission, the propagation of the EUV wave, and the relation between both phenomena. Observations are reported in Section~2. An analysis of the data and results are in Section~3, and a brief discussion and conclusions are provided in Section~4. 

\section{Observations}

On 10 May 2022 at 13:50~UT,  an X~1.5 class solar flare occurred originating from active region NOAA AR13006. The solar flare was accompanied by a shock event. This event began less than 10~min after a C--class flare from the eastern limb of the Sun. 

\subsection{Radio dynamic spectrum}

I-LOFAR was conducting solar observations on 10 May 2022 with data recorded by the REAL-time Transient Acquisition (REALTA) system \citep{Murphy2021}. The I-LOFAR station forms part of the International LOFAR Telescope \citep{vanHaarlem2013} and observes with low band antennas at 10--90~MHz, and high band antennas from 110--230~MHz. The telescope can be configured to observe a target with both antenna sets, utilising the full frequency band of 10--240~MHz. Using the REALTA pipeline allows observations to be recorded at 5.12~$\mu$s time resolution, and 195~kHz frequency resolution. The data for this project were stored at a down-sampled time resolution of 1.2~ms. I-LOFAR captured a dynamic spectrum with a number of Type III bursts prior to the X-ray flare observed by GOES. A Type II radio burst was recorded 3~min after the peak of the X-ray emission, lasting for approximately 15~min (see Figure \ref{fig:1_dynspec_euv}). The Type II burst displayed a band split in both the fundamental and harmonic bands, as seen in previous work by \cite{Maguire2020} and \cite{Magdalenic2020}. 

ORFEES was also observing the Sun during the X--class event at a cadence of 0.1~s and a frequency range of 144--1000~MHz. At the peak of the flare (13:57~UT), a high-frequency Type III radio burst was detected between 300--800~MHz. 

During the Type II burst observed by I-LOFAR, a Type IV burst also occurred at frequencies ranging from $\sim$100 to $\sim$500~MHz. This diffuse broadband radio emission was observed by all three radio instruments, I-LOFAR, ORFEES,  and NRH.

\subsection{AIA EUV imaging}

The on--disk signatures of the global EUV wave were identified using observations from the AIA on board SDO. The global EUV wave could be visually identified and compared to the position of the radio emission using three--colour running ratio images \citep[as shown in Figure~\ref{fig:1_dynspec_euv}; see][for more details]{Downs:2012}. As can be seen in Figure~\ref{fig:1_dynspec_euv}, the wave front was highly directional, mainly propagating to the north-east of the erupting active region, with some indication of wave-front propagation to the south of the active region, similar to the observed radio emission. The kinematics of the global wave front were estimated using the arc-sector approach of \citet{Long:2014} to enable a direct comparison with the kinematics of the observed radio emission. A series of arc sectors of 10$^{\circ}$ width were applied to the 12~s cadence data from the AIA 211~\AA\ pass band to produce distance-time plots that showed the temporal evolution of the wave front along each arc sector. The leading edge of the wave front could then be manually identified as previously described by \citet{Long:2021} and compared to the observed radio signature.

\subsection{Radio imaging}

Radio images sampled at 0.25~s cadence at 154~MHz and 408~MHz were obtained from NRH (Figure~\ref{fig:1_dynspec_euv}(b)). This provided spatial information for the location of the radio emission that is accurate to $10"$ at 154~MHz. The location of the peak radio flux was measured for each image, and the motion of the radio source was recorded for comparison to the EUV wave propagation.
The NRH radio images were used for timing analysis to confirm that the Type II radio burst captured with I-LOFAR originated from the flaring active region.

Two clear 154~MHz radio sources on the solar disk were observed propagating away from the active region for a short time after the flare. The southward propagating radio emission quickly faded after approximately 40~s. The northward propagating radio emission was observed for approximately 7~min, with clear motion of the peak along the axis of propagation of the EUV wave. The motion of the 154~MHz peak was measured to determine the velocity across the solar disk.

The radio imaging also revealed that a Type III noise storm observed prior to the X--class flare. The source originated from AR3007 on the eastern limb where an earlier C--class flare originated and was not related to the X--class event.

\section{Data analysis and results}

\subsection{Dynamic spectrum}

Background subtraction was performed on the dynamic spectrum to flatten the frequency response of the LOFAR antennas. The Type II burst within the spectrum displayed a number of fine structures resolvable at high time resolution. The fundamental (start: 60~MHz, $f_p$) and harmonic (start: 120~MHz, $2f_p$) bands are observed with a clear separation in Figure \ref{fig:1_dynspec_euv}(a). Within these bands, herringbone structures are visible \citep{carley2013, morosan2019}. The fundamental band displays band-splitting. Using the same method as described in \cite{vrsnak2002} and \cite{Maguire2020}, an estimate of the Alfv\'en Mach number was made for the duration of the Type II burst from the separation of the band-splitting. The Alfv\'en Mach number ranged between 1.5 and 1.6 throughout the shock. This burst had a drift rate of 0.59~MHz s$^{-1}$.

Points were selected along the leading edge of the Type II burst fundamental band at regular time intervals. The Mann solar density model \citep{mann2005} was used to calculate the shock speed from the leading edge of the Type II burst. Other electron density models were also investigated, including densities typical of both coronal hole and quiet-Sun conditions \citep{Newkirk1961, doyle1999}; however, the event studied here occurred near an active region so these were not suitable. A mean shock velocity of $420^{+150}_{-120}$~km~s$^{-1}$ was determined using this method (see Figure \ref{fig:1_dynspec_euv}).

Other fine structures were observed within the event, including flag, spike, and sail-like features consistent with those observed in \cite{Magdalenic2020}. An interesting phenomenon observed at the end of the Type II burst (approx. 14:15~UT) is the appearance of two further lanes or bands that mirror the end of the Type II burst but at a later time. These later bands also appear to display band-splitting but their source is unknown.

A Type IV radio burst was recorded after the Type II burst, ranging in frequencies from 20--500~MHz. The Type IV burst had fine vertical structures resembling herringbone structures.

\subsection{EUV wave velocity}

The velocity of the EUV wave front was measured accounting for the propagation across the solar disk. Two methods were used to estimate this velocity, CorPITA \citep{Long:2014} and Wavetrack, \citep{stepanyuk2022}. Figure \ref{fig:1_dynspec_euv} \textit{(b)} contains the morphology of the EUV wave front for the duration of the event. 
The area surrounding the active region was divided into $10^{\circ}$ segments. A velocity profile for each segment was created using wave-front detection with CorPITA. Segments with angles of propagation matching the 154~MHz radio emission were then examined in greater detail to determine the acceleration and dispersion profiles.
The EUV wave velocity was measured for the region spatially overlapping the 154~MHz radio emission. The mean velocity of the wave during this time was $\sim$~500~km~s$^{-1}$.

A second method was also employed to estimate the EUV wave velocity, using \textit{Wavetrack}\footnote{https://gitlab.com/iahelio/mosaiics/wavetrack}, a wavelet decomposition technique for wave-front detection and analysis. The Wavetrack object masks were applied to the running difference, using the AIA 193 \AA\ observations. The Wavetrack was able to capture the extent of the wave in the consecutive time steps very well.
To study the kinematics of the coronal front features in detail based on the Wavetrack output, the Fourier local correlation tracking (FLCT) method was employed \citep{Welsch2004, Fisher2008}, followed by centre of the mass velocity calculation. FLCT calculations show velocities of $\sim$~700~km~s$^{-1}$, which is at the upper end of the error for the more commonly used CorPITA method.

\subsection{Radio imaging} 

\begin{figure}
   \centering
   \includegraphics[width=0.49\textwidth]{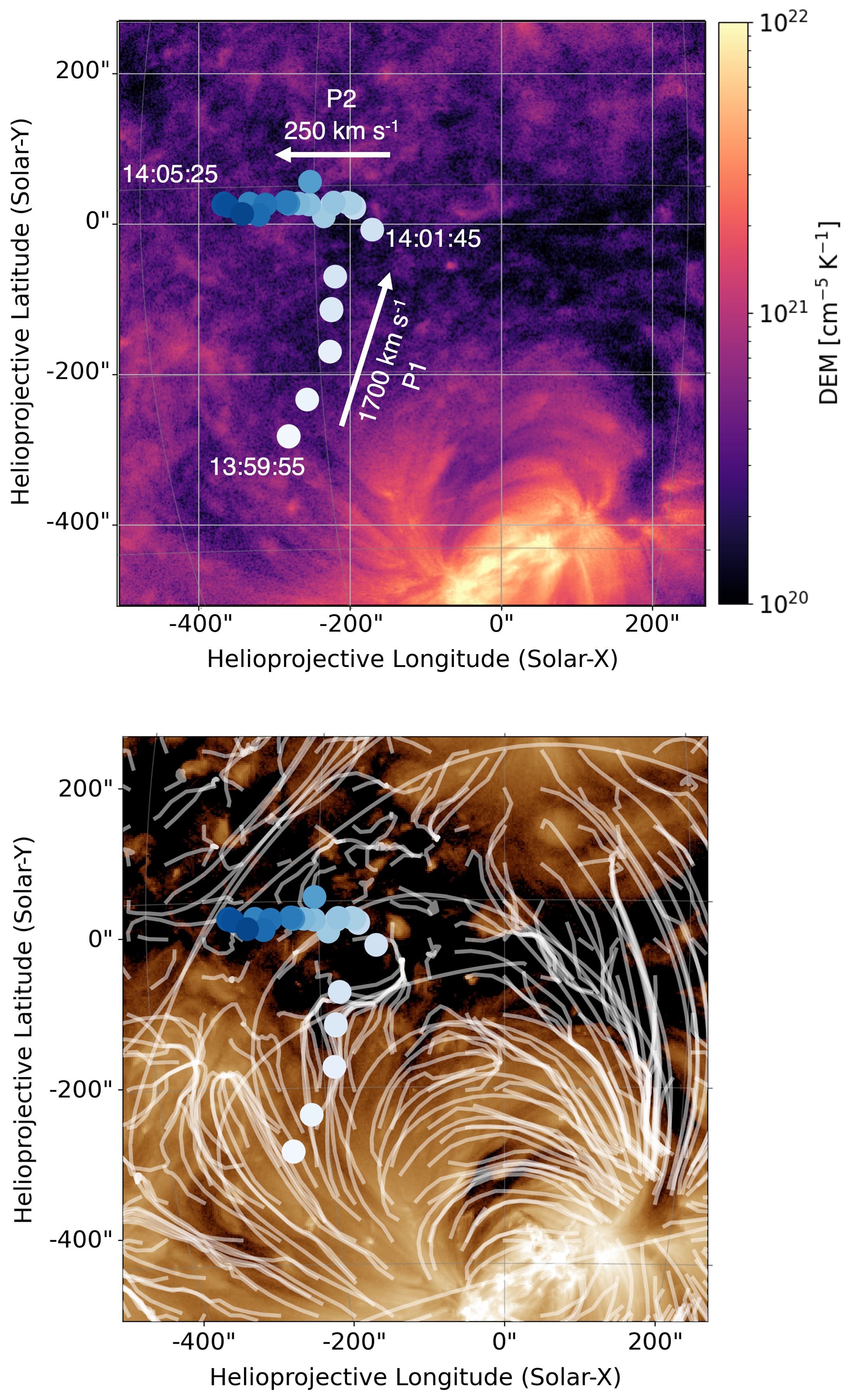}
       \caption{ Path of the 154 MHz radio source. (a): Line-of-sight-integrated DEM map. The white-to-blue points show the time evolution of the 154~MHz radio source in 10~s intervals. The velocities of the points in Phase 1 and Phase 2, obtained from 0.25s data, are indicated by the labelled arrows. (b): Combination of an AIA 193~Å image during the event with a PFSS extrapolation. The radio emission clearly moves around a higher Alfv\'en speed region, favouring lower magnetic field strength regions.}
   \label{fig:2_radio_path}
\end{figure}

\begin{figure}
   \centering
   \includegraphics[width=0.49\textwidth]{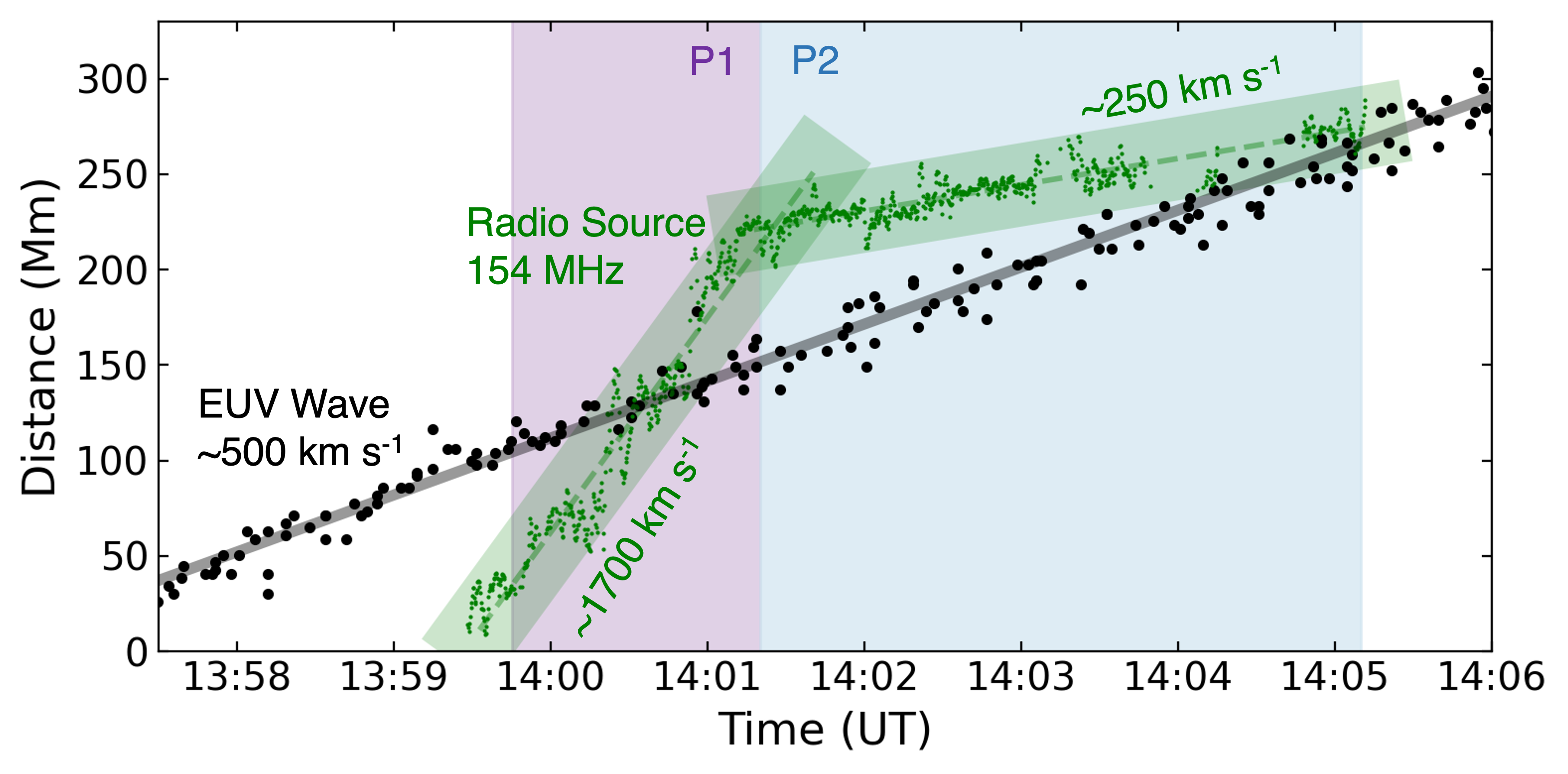}
       \caption{Distance-time plot for CorPITA EUV wave (black points) and 0.25~s time resolution radio source (green points) with errors for each highlighted by shaded regions. The purple region (P1) highlights the region when the radio source initially travelled at a higher velocity, later slowing in the blue-shaded region (P2) once it reached the region of lower density. These correspond to the same emission marked in Figure \ref{fig:2_radio_path}. The velocity of the EUV wave was 500~km~s$^{-1}$ and the radio source velocities were 1700~km~s$^{-1}$ and 250~km~s$^{-1}$.}
   \label{fig:3_dist_time}
\end{figure}

NRH images at 154~MHz and 408~MHz were examined to determine the emission propagation characteristics. \textit{SolarSoft} \citep{Freeland1998} was used to produce images from the Nançay observations using the default pipeline settings. The location of the peak radio emission from the 154~MHz NRH imaging on the solar disk was recorded for each 0.25~s timestamp. The peak emission was observed to drift upwards across the solar disk contemporaneously with the propagation of the EUV wave in the low solar corona (Figure~\ref{fig:2_radio_path}). In order to determine if a correlation existed between the motion of the two events, the movement of the peak radio emission was recorded across the Sun. The 154~MHz band displayed rapidly moving north- and southward peaks. 
Two distinct velocity profiles can be seen in Figure~\ref{fig:3_dist_time} initially travelling at $\sim1750$~km~s$^{-1}$ in Phase 1, then slowing to $\sim230$~km~s$^{-1}$ in Phase 2. The abrupt change in speed corresponds to the change in direction of the radio emission source from 14:01:55 UT, as highlighted in Figure~\ref{fig:2_radio_path}. The change in velocity of the radio source corresponds to the shock entering a region of lower density and Alfv\'en speed. This motion is also observed in Figure~\ref{fig:1_dynspec_euv} where the radio source appears to overtake the EUV wave front. Some uncertainty remains in the radio velocity estimates due to projection effects observing emission on-disk and radio source height estimates of 1.2~R$_{\odot}$ for the 154 MHz emission.

Magnetic field values were calculated using a potential field source surface (PFSS) extrapolation at 1.2~R$_{\odot}$ from Global Oscillation Network Group (GONG) magnetograms. Densities were determined from differential emission measure (DEM) maps (see \citet{Stansby2020}) and \texttt{demregpy}\footnote{https://github.com/alasdairwilson/demregpy}. The DEM map was integrated across all AIA temperature ranges $5.6 \leq log_{10}T \leq 6.8$~K to produce a line-of-sight DEM. Regarding the DEM initial conditions, full details on the inversion method, in addition to assumptions relating errors and temperature resolution can be found in \citet{hannah2012} where these authors tested the code on both simulated and real Hinode/EIS and SDA/AIA data.  A height of 1.2~R$_{\odot}$ was used to determine the magnetic field values using the Mann density model. These values could then be used to create an Alfv\'en speed map of the region through which the radio source travelled. The Alfv\'en speed dropped from $\sim550$~km~s$^{-1}$ near the active region to $\sim$~300 km~s$^{-1}$ in the quiet-Sun region. The cumulative distance versus time plot in Figure \ref{fig:3_dist_time} highlights the rapid onset acceleration and gradual decay in velocity as the radio source propagates upwards across the solar disk. 

Stationary radio emission was observed at 408~MHz for the duration of the flare and CME event directly over the active region, and it occurred for the same duration and frequency range within the I-LOFAR dynamic spectrum as the Type IV radio burst.

\subsection{Coronal channel}

The 154~MHz radio source tracking the EUV wave was observed to deviate from a great circle path significantly. When compared to the Alfv\'en map, the radio source appears to rapidly move away from the higher Alfv\'en speed of the active region into a region of lower Alfv\'en speed. A physical explanation for this observed path could be the presence of a wave guide that formed by the surrounding conditions in the corona, creating a region favourable for electron acceleration in the propagating shock front. Understanding the complete picture of shock formation and evolution during the event is difficult due to the complex magnetic environment surrounding the active region. However, the EUV wave and radio emission propagate into a region of lower magnetic field strength, while the electron density does not vary dramatically. A high Alfv\'en speed suggests a shock wave is being formed for the majority of the duration of the event.

The EUV wave front is not noticeably impacted by the coronal channel. It is possible that conditions conducive to radio emission in the upper corona are generated at different points along the shock front during the propagation. This would suggest that the observed radio emission is occurring where the shock is propagating quasi-perpendicular to the local magnetic field, generating the observed apparent non-uniform propagation. A similar phenomenon was observed in \cite{koukras2020}. Furthermore, electron density variations, while estimated with DEMs, are difficult to constrain at high spatial resolution at different heights in the corona. It is therefore also possible that the radio emission is occurring at different heights along the expanding shock front, further complicating the calculation of the velocity and path taken. Examining the direction of the magnetic field was beyond the scope of this research.

\section{Discussion and conclusions}

Multi-wavelength solar observations with NRH, AIA, and I-LOFAR on 10 May 2022 revealed an EUV wave and radio emission associated with an X-class flare and shock wave. Analysis of the observations from each instrument appears to show a spatial and temporal correlation between the EUV wave and radio emission imaged at the shock front. The EUV wave had a constant velocity of ($\sim$~500~km~s$^{-1}$). The associated 154~MHz radio source was measured to have an initial horizontal velocity of 1700~km~s$^{-1}$, changing to 250~km~s$^{-1}$. Using the Mann density model, a radial velocity of $\sim$~420~km~s$^{-1}$ was derived from the Type II burst, lying within the range of the EUV wave velocity. Our observations allowed us to examine shock propagation at different heights throughout the solar corona as the shock propagated across the solar disk. 

This event was captured during the initial formation, propagation, and expansion of a coronal shock front. This may explain the lack of correlation between the EUV wave and 154~MHz radio imaging prior to 14:00~UT, as the shock was not self-similar in the initial expansion stage \citep{subramanian2014, uralov2019}. The event can be compared with emission observed from an off-limb event in 2005 in which an EUV wave, Type II radio burst, and low-frequency NRH imaging were observed from a CME event \citep{vrsnak2005}. The EUV wave was observed to slow and disperse into a region of lower density; this has been observed during previous events and predicted in simulations \citep{wu2001, veronig2006}. 

The most probable scenario arising from all emission moving at similar velocities is that the emission source is linked. In this case the observed propagation outwards from the active region that produced the X--class flare suggests a shock wave that accelerated electrons in the upper corona, producing radio emission, while exciting  plasma in the low corona at the base of the shock forming an EUV wave. The physical mechanism producing the 154~MHz radio emission and dynamic spectrum is most likely shock-accelerated electrons (plasma emission). The faint, weak EUV wave indicates low levels of disruption to the low solar corona, which propagates across the disk below the radio emission. Models of these events indicate regions where radio emission is expected to be produced in regions where the shock is propagating normally to the local magnetic fields \citep{zucca2014, morosan2019}.

Finally, we observed the motion of radio emission travelling along a coronal channel, bounded by regions of higher magnetic field strength. Phenomena such as EUV wave reflection and refraction support the case that shocks can be diverted and directed by conditions in the low corona. Two possible phenomena are occurring during this event, as interpreted from the motion of the 154 MHz radio emission. Either the shock is being guided through the channel of lower density, or the radio source is visible at points where the shock is quasi-perpendicular to the local magnetic field. The tracking of the motion of low-frequency radio emission also suggests that this guiding behaviour can occur at higher regions of the outer corona. Future analysis of similar events could probe the local motion of an EUV wave in conditions where shock propagation travels through channels of lower magnetic field strengths, as well as search for a similar motion at different radio frequencies at different heights of a shock in the corona.

\begin{acknowledgements}
I-LOFAR receives funding from Science Foundation Ireland (SFI), the Department of Further and Higher Education, Research, Innovation and Science (DFHERIS) and the Department for Communities of the N. Ireland Executive via AOP. REALTA was funded by SFI and Breakthrough Listen. JGD would like to thank the Leverhulme Trust for an Emeritus Fellowship. DML is grateful to the Science Technology and Facilities Council for the award of an Ernest Rutherford Fellowship (ST/R003246/1). This work is partly supported by the European Union’s Horizon 2020 research and innovation program under grant agreement 952439, project STELLAR (Scientific and Technological Excellence by Leveraging LOFAR Advancements in Radio Astronomy). K.K. and O.S. acknowledge support from the Bulgarian National Science Fund, VIHREN program, under contract KP-06-DV-8/18.12.2019. 
\end{acknowledgements}

\bibliographystyle{aa} 
\bibliography{references}

\end{document}